\documentclass[journal, final]{IEEEtran}

\usepackage{cite}      
\usepackage{graphicx}  
\usepackage{subfigure} 
\usepackage{url}       
\usepackage{stfloats}  
\usepackage{amsmath}   
\usepackage{amssymb}   

\usepackage{amsmath}
\usepackage{amsthm}
\usepackage{amssymb}
\usepackage{cite}
\usepackage{graphicx}
\usepackage{booktabs}
\usepackage{subfigure}

\newcommand{\ra}{\rightarrow}

\newcommand{\hf}{\frac{1}{2}}

\newcommand{\q}{\mathsf{q}}

\newtheorem{theo}{Theorem}
\newtheorem{corollary}{Corollary}

\begin{document}
%
\title{Capacity of the Gaussian Relay Channel with Correlated Noises to Within a Constant Gap
}


\author{ Lei Zhou, {\it Student Member, IEEE} and
	Wei Yu, {\it Senior Member, IEEE}}

\maketitle

\begin{abstract}
This paper studies the relaying strategies and the approximate capacity of the classic three-node Gaussian relay channel, but where the noises at the relay and at the destination are correlated. It is shown that the capacity of such a relay channel can be achieved to within a constant gap of $\hf \log_2 3 =0.7925$ bits using a modified version of the noisy network coding strategy, where the quantization level at the relay is set in a correlation dependent way. As a corollary, this result establishes that the conventional compress-and-forward scheme also achieves to within a constant gap to the capacity. In contrast, the decode-and-forward and the single-tap amplify-and-forward relaying strategies can have an infinite gap to capacity in the regime where the noises at the relay and at the destination are highly correlated, and the gain of the relay-to-destination link goes to infinity.
\end{abstract}

\section{Introduction}
The relay channel models a communication scenario where an
intermediate relay is deployed to assist the direct communication between a source and the destination.  Although the capacity of the relay channel is still not known exactly even for the Gaussian case, much progress has been made recently in the characterization of its approximate capacity \cite{Avestimehr_relaynetwork, Chang_constantbitrelay, Kim_noisy_network_coding}.

In the classic Gaussian relay channel, the noises at the relay and at the destination are independent. In many practical systems, however, the noises at the relay and at the destination may be {\em correlated}. This may arise, for example, due to the presence of a common interference, which in a practical system is often treated as a part of the background noise, but nevertheless contributes to the correlation between the noises.

The Gaussian relay channel with correlated noises has been studied
in \cite{Cui_correlated}, where relaying strategies such as the
decode-and-forward and the compress-and-forward schemes are studied
in full-duplex or half-duplex modes. 
Likewise, the effect of noise correlation for the single-tap amplify-and-forward
scheme has been studied for the diamond network and the two-hop
parallel relay network in \cite{Jafar_correlated}. 
In both papers, noise correlation has been found to be beneficial. Neither \cite{Cui_correlated} nor  \cite{Jafar_correlated}, however, addresses the question of whether the classic relaying strategies are able to achieve to within constant bits of the capacity for the relay channel with correlated noises.


Inspired by the recent work \cite{Avestimehr_relaynetwork} and \cite{Kim_noisy_network_coding}, where the quantize-map-and-forward and the noisy network coding strategies with fixed quantization level at the relays are shown to achieve the capacity of arbitrary Gaussian relay networks with uncorrelated noises to within a constant gap, this paper shows that such strategies are also capable of approximating the capacity of the three-node Gaussian relay channel with correlated noises. However, unlike the existing schemes of \cite{Avestimehr_relaynetwork} and \cite{Kim_noisy_network_coding}, this paper shows that the relay quantization level needs to be modified to be noise-correlation dependent in the correlated-noise case. As a corollary, this paper also establishes that the conventional compress-and-forward scheme \cite{Cover1979} achieves to within constant bits of the capacity for the Gaussian relay channel in the correlated-noise case as well. Finally, in contrast to the case with uncorrelated noises, the decode-and-forward and the single-tap amplify-and-forward strategies can have an infinite gap to capacity, when the noise correlation goes to $\pm 1$ and the gain of the relay-to-destination link goes to infinity.

\section{Channel Model}

\begin{figure} [t]
\centering
\includegraphics[width=2.7in]{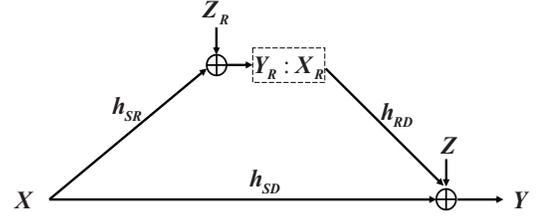}
\caption{Three-node Gaussian relay channel with correlated noises}
\label{channel_model}
\end{figure}

This paper considers a real-valued discrete-time three-node Gaussian relay channel as depicted in Fig.~\ref{channel_model}, which consists of a source $X$, a destination $Y$, and a relay. The relay observes a noise-corrupted version of the source signal, denoted by $Y_R$, and transmits $X_R$ to the destination. The source-to-destination channel is denoted $h_{SD}$, the  relay-to-destination channel $h_{RD}$, and the source-to-relay
channel $h_{SR}$. The additive Gaussian noises at the relay and at the destination are denoted as $Z_R$ and $Z$ respectively.
Mathematically, the channel model is:
\begin{eqnarray}
Y_R &=& h_{SR}X + Z_R,\\
Y &=& h_{SD}X + h_{RD}X_R + Z.
\end{eqnarray}
Without loss of generality, the power constraints at the source and at
the relay can both be normalized to one, i.e., $\mathbb{E}[X^2] \le
1$ and $\mathbb{E}[X_R^2] \le 1$, and so can the noise variances, i.e.,
$Z_R\sim \mathcal{N}(0, 1)$ and $Z \sim \mathcal{N}(0, 1)$. Different
from most of the literature that assumes independence between $Z_R$
and $Z$, this paper introduces a correlation between the two noises
\begin{equation}
\rho_z \triangleq \frac{\mathbb{E}\left[Z_R
Z\right]}{\sqrt{\mathbb{E}[|Z_R|^2]\mathbb{E}[|Z |^2]}}.
\end{equation}
Note that $Z$ and $Z_R$ are both i.i.d.\ in time. Further, the relay operation is causal. 

\section{Within Constant Bits of the Capacity} \label{Theorem_section}


To approach capacity, the relaying strategy must take advantage of the noise correlation. 
Consider the limiting scenario of $\rho_z \ra \pm 1$. The relay's observation becomes more
and more useful to the destination in this case, thus an increasingly fine
quantization resolution at the relay is required --- the fixed
quantization strategy of \cite{Avestimehr_relaynetwork} and \cite{Kim_noisy_network_coding} would result in significant inefficiency.  The main contribution of this paper is to introduce a {\em correlation-aware} quantization strategy at the relay, which better exploits the noise correlation and achieves to within $\frac{1}{2} \log_2 3$ bits of the capacity of the Gaussian relay channel with correlated noises.

\begin{theo} \label{constant_gap_joint_decoding}
The capacity of the three-node Gaussian relay channel with correlated
noises, as shown in Fig.~\ref{channel_model}, can be achieved to
within $\frac{1}{2}\log_2 3$ bits to capacity using a noisy network
coding strategy
with independent Gaussian inputs $X \sim \mathcal{N}(0, 1)$,
$X_R \sim \mathcal{N}(0, 1)$ and Gaussian quantization at
the relay with quantization variance $\q^* = 2(1 - \rho_z^2)$.
\end{theo}
\begin{IEEEproof}
First, the capacity of the relay channel is upper bounded by
the cut-set bound, i.e.,
\begin{eqnarray}
\overline{C} & = &\max_{p(x, x_R)} \min \{I(X,X_R; Y), I(X; Y, Y_R|X_R)
\} \nonumber \\
&=& \max_{\rho_x}\min  \left\{  \hf \log(1 + h_{SD}^2 + h_{RD}^2 + 2\rho_x h_{SD}h_{RD}), \right. \nonumber \\
&& \left. \hf \log\left( 1+ \frac{(1-\rho_x^2)(h_{SD}^2 + h_{SR}^2 - 2\rho_z h_{SD}h_{SR})}{1 - \rho_z^2} \right) \right\} \nonumber \\
&\le& \min \left\{ \hf \log(1 + h_{SD}^2 + h_{RD}^2 + 2h_{SD}h_{RD}), \right. \nonumber \\
&& \quad \quad \;\; \left. \hf \log  \left(1+ \frac{h_{SD}^2 + h_{SR}^2 - 2\rho_z h_{SD}h_{SR}}{1 - \rho_z^2} \right) \right\}  \nonumber \\
&=& \min\{R_{UB1}, R_{UB2} \}, \label{cut-set-bound}
\end{eqnarray}
where $\rho_x$ is the correlation between $X$ and $X_R$.

The achievable rate by noisy network coding or compress-and-forward
with joint decoding can be readily obtained from
\cite[Proposition~2]{Dabora_jointdecoding} and
\cite[Theorem~1]{Kim_noisy_network_coding}:
\begin{eqnarray}
R &=&\min \{I(X,X_R; Y)-I(Y_R;\hat{Y}_R|X, X_R, Y) , \nonumber \\
&& \quad \quad \; I(X; Y, \hat{Y}_R|X_R)\} \nonumber \\
&=& \min \{ R_1, R_2 \} \label{rate:jointdecoding}
\end{eqnarray}
for any distribution
\begin{equation}
p(x, x_R, y_R, \hat{y}_R) = p(x)p(x_R)p(y_R|x, x_R)p(\hat{y}_R|x_R, y_R). \nonumber
\end{equation}
Substitute independent Gaussian distributions $X \sim \mathcal{N}(0, 1)$
and $X_R\sim \mathcal{N}(0, 1)$ into (\ref{rate:jointdecoding}), and set
$\hat{Y}_R = Y_R + e$, where the quantization noise $e \sim \mathcal{N}(0, \q)$
is independent with everything else, we have
\begin{eqnarray}
\lefteqn{R_1 = I(X,X_R; Y)-I(Y_R;\hat{Y}_R|X, X_R, Y)} \nonumber \\
&=& \hf \log(1 + h_{SD}^2 + h_{RD}^2) - \hf \log \left(1 +
\frac{1-\rho_z^2}{\q} \right),  \label{R1}
\end{eqnarray}
and
\begin{eqnarray}
R_2&=& I(X; Y, \hat{Y}_R|X_R) \nonumber \\
&\stackrel{(a)}{=}& \hf \log(1 + h_{SD}^2)  \nonumber \\
&&+ \hf \log \left(\frac{\q + \sigma^2_{h_{SR}X + Z_R | h_{SD}X+Z}}{\q + 1-\rho_z^2} \right) \nonumber \\
&=& \hf \log \left( 1 + \frac{\q +(\q +1)h_{SD}^2 + h_{SR}^2 - 2\rho_z h_{SD}h_{SR}}{1 - \rho_z^2} \right) \nonumber \\
&&- \hf \log \left( 1 + \frac{\q}{1- \rho_z^2} \right) \label{R2},
\end{eqnarray}
where in $(a)$ the conditional variance of $h_{SR}X +
Z_R $ given $ h_{SD}X+Z$ is calculated as
\begin{eqnarray}
\lefteqn{\sigma^2_{h_{SR}X + Z_R | h_{SD}X+Z}} \nonumber \\
&=& \mathbb{E}[|h_{SR}X + Z_R|^2] - \frac{|\mathbb{E}[(h_{SR}X + Z_R)
(h_{SD}X+Z)]|^2}{\mathbb{E}[|h_{SD}X+Z|^2]} \nonumber \\
&=& \frac{1 - \rho_z^2 + h_{SR}^2 + h_{SD}^2 -2\rho_z h_{SR}h_{SD}}{1 +
h_{SD}^2}.
\end{eqnarray}
Comparing $R_1$ and the upper bound $R_{UB1}$, we have
\begin{eqnarray}
\lefteqn{R_{UB1} - R_1} \nonumber \\
&=& \hf \log(1 + h_{SD}^2 + h_{RD}^2 + 2h_{SD}h_{RD}) \nonumber \\
&& - \hf \log(1 + h_{SD}^2 + h_{RD}^2) + \hf \log \left(1 + \frac{1-\rho_z^2}{\q} \right) \nonumber \\
&=& \hf \log \left(\frac{1 + h_{SD}^2 + h_{RD}^2 + 2h_{SD}h_{RD}}{2 + 2h_{SD}^2 + 2h_{RD}^2} \right)  \nonumber \\
&&+ \hf \log \left(1 + \frac{1-\rho_z^2}{\q} \right) + \hf \nonumber \\
&<& \hf \log \left(1 + \frac{1-\rho_z^2}{\q} \right) + \hf. \label{gap1}
\end{eqnarray}
Comparing $R_2$ and the upper bound $R_{UB2}$, we have
\begin{eqnarray}
\lefteqn{R_{UB2} - R_2} \nonumber \\
&=& \hf \log\left(1 + \frac{h_{SD}^2 + h_{SR}^2 - 2\rho_z h_{SD}h_{SR}}{1 - \rho_z^2} \right)  \nonumber \\
&& -\hf \log \left(1 + \frac{\q +(\q +1)h_{SD}^2 + h_{SR}^2 - 2\rho_z h_{SD}h_{SR}}{1 - \rho_z^2} \right) \nonumber \\
&& + \hf \log \left( 1+ \frac{\q}{1- \rho_z^2} \right) \nonumber \\
&<& \hf \log \left( 1+ \frac{\q}{1- \rho_z^2} \right). \label{gap2}
\end{eqnarray}
The gap between the cut-set bound $\overline{C}$ and the achievable
rate $R$ is then upper bounded by the maximum of
(\ref{gap1}) and (\ref{gap2}), i.e.
\begin{eqnarray}
\overline{C} - R &\le& \min\{R_{UB1}, R_{UB2}\} - \min \{R_1, R_2 \} \nonumber \\
&\le& \max \{R_{UB1} - R_1, R_{UB2} - R_2 \} \nonumber \\
&<& \max \left\{\hf \log \left(1+ \frac{1-\rho_z^2}{\q} \right) + \hf,
\right. \nonumber \\
&& \quad \quad \;\; \left.\hf \log \left(1 + \frac{\q}{1- \rho_z^2} \right)
\right\}. \label{gap}
\end{eqnarray}
The first term above monotonically decreases with $\q$, while the second term monotonically increases with $\q$. To minimize the maximum of the two terms, we set
\begin{eqnarray}
\hf \log \left(1 + \frac{1-\rho_z^2}{\q^*} \right) + \hf = \hf \log \left( 1+ \frac{\q^*}{1- \rho_z^2} \right),
\end{eqnarray}
which results in $\q^* = 2(1- \rho_z^2)$. Substituting $\q^*$ into
(\ref{gap}), we have $\overline{C} - R < \hf \log_2 3  =0.7925$.
\end{IEEEproof}

\vspace{0.5em}
In addition, it can be shown that the conventional compress-and-forward rate is also within the same constant gap to capacity. To prove this directly would have been quite involved (see \cite{Chang_constantbitrelay} for the computation of the gap for the case of $\rho_z=0$). Instead, we obtain the result as a direct consequence of Theorem~\ref{constant_gap_joint_decoding}.

\begin{corollary} \label{constant_gap_sequantial_decoding}
The following rate, which is achieved by the classic compress-and-forward
strategy on the three-node Gaussian relay channel with correlated
noises shown in Fig.~\ref{channel_model}:
\begin{equation} \label{CF_rate}
R_{CF} = \hf \log\left(1 + h_{SD}^2 + \frac{(h_{SR} - \rho_z h_{SD})^2}{1 -
\rho_z^2 + \q_{c}}\right),
\end{equation}
where
\begin{equation} \label{q_c}
\q_c = \frac{(1 - \rho_z^2)(1 + h_{SD}^2) + (h_{SR} - \rho_z h_{SD})^2
}{h_{RD}^2}
\end{equation}
is within $\frac{1}{2}\log_2 3$ bits to the capacity.
\end{corollary}

\begin{IEEEproof}
The rate expression $R_{CF}$ for the correlated-noise Gaussian relay
channel has been obtained in \cite[Proposition~5]{Cui_correlated}.
The derivation is based on the classic
compress-and-forward rate for the relay channel by Cover and El Gamal
\cite[Theorem~6]{Cover1979}, which is
$R_{CF} = I(X; \hat{Y}_R, Y |X_R)$
subject to $I(X_R; Y) \ge I(Y_R; \hat{Y}_R |X_R, Y)$ for some joint
distribution
$p(x)p(x_R)p(y_R|x, x_R)p(\hat{y}_R|x_R, y_R)$.
Using the same signaling scheme as in
Theorem~\ref{constant_gap_joint_decoding}, i.e.,
$X \sim \mathcal{N}(0,1)$ and $X_R \sim \mathcal{N}(0,1)$  are independent,
and $\hat{Y}_R = Y_R + e$, where $e \sim \mathcal{N}(0, \q_c)$ is chosen
to satisfy the relay-destination rate constraint, we obtain
(\ref{CF_rate}).

In the following, we prove the constant gap result for the compress-and-forward rate by showing that $R_{CF}$ in (\ref{CF_rate}) is greater than the noisy network coding rate, i.e., $R_{CF} \ge \min(R_1, R_2)$, where $R_1$ and $R_2$ are as in (\ref{R1}) and (\ref{R2}) respectively. Substituting $\q_c$ in (\ref{q_c}) as $\q$ in $R_1$ and $R_2$, it is easy to verify that $R_1(\q_c) = R_2(\q_c) = R_{CF}$. Since $R_1$ increases with $\q$ and $R_2$ decreases with $\q$, we have $R_{CF}= \min\{R_1(\q_c), R_2(\q_c)\} = \max_{\q} \min\{R_1(\q), R_2(\q)\}  \ge \min\{R_1(\q^*), R_2(\q^*)\}$ for any $\q*$ and in particular for $\q^*=2(1-\rho_z^2)$. Since it has been show in Theorem~\ref{constant_gap_joint_decoding} that $\min\{R_1(\q^*),  R_2(\q^*)\}$ is within $\hf \log 3$ bits of the cut-set upper bound, so is $R_{CF}$.
\end{IEEEproof}

\section{Suboptimality of Decode-and-Forward and Single-Tap Amplify-and-Forward }


The decode-and-forward and the single-tap amplify-and-forward
strategies have been shown to achieve to within a constant gap to the capacity
of the Gaussian relay channel with uncorrelated noises
\cite{Avestimehr_relaynetwork, Chang_constantbitrelay}. In this
section, we show that this is no longer the case when noises are
correlated.  


\subsection{Decode-and-Forward}

Consider a decode-and-forward strategy as described in
\cite[Appendix~A]{Avestimehr_relaynetwork}, in which when the
source-to-relay link is weaker than the source-to-destination link,
i.e., $h_{SR} \le h_{SD}$, the relay is simply ignored, otherwise the
relay decodes and forwards a bin index to the destination as
in the original scheme of \cite{Cover1979}.
The following rate is achievable:
\begin{multline} \label{DF_rate}
R_{DF} = \max \left\{  \hf \log(1 + h_{SD}^2), \right.\\
	\left. \min \left\{  \hf \log(1 +h_{SR}^2),
	\hf \log(1 +h_{SD}^2 + h_{RD}^2) \right\} \right\}
\end{multline}
In the extreme scenario where $\rho_z = 1$ and
\begin{equation} \label{extreme_scenario}
h_{RD}^2 \gg h_{SR}^2 \gg h_{SD}^2 \gg 1,
\end{equation}
the above decode-and-forward rate (\ref{DF_rate}) becomes
\begin{eqnarray} \label{RDF}
R_{DF} = \hf \log(1 + h_{SR}^2).
\end{eqnarray}
Meanwhile, when $\rho_z =1$, the cut-set bound
(\ref{cut-set-bound}) becomes
\begin{equation}\label{DF_bound}
\overline{C} = \hf \log(1 + h_{SD}^2 + h_{RD}^2 + 2h_{SD}h_{RD}).
\end{equation}
Comparing (\ref{RDF}) with (\ref{DF_bound}), we observe that
\begin{eqnarray}
\overline{C} - R_{DF} &=& \hf \log \left(\frac{1 + h_{SD}^2 + h_{RD}^2 + 2h_{SD}h_{RD}}{1 + h_{SR}^2} \right)  \nonumber \\
&\rightarrow& \hf \log \left(\frac{h_{RD}^2}{h_{SR}^2} \right),
\end{eqnarray}
which is unbounded in the asymptotic regime (\ref{extreme_scenario}).
This is not unexpected, because the decoding at the relay eliminates
the noise.  Therefore, noise correlation is not exploited.


\subsection{Single-Tap Amplify-and-Forward}


In the single-tap amplify-and-forward, the relay scales the current
observation and forwards to the destination in the next time instance, i.e.,
\begin{equation}
X_R[i] = \alpha(h_{SR}X[i-1] + Z_R[i-1]),
\end{equation}
where $\alpha \le \frac{1}{\sqrt{1 + h_{SR}^2}}$ is chosen to satisfy the
power constraint at the relay. 
Since $Y[i] = h_{SD}X[i] + h_{RD}X_R[i] + Z[i]$,
the relay channel is now converted into a single-tap
inter-symbol-interference (ISI) channel:
\begin{equation}
Y[i] = h_{SD}X[i] + \alpha h_{RD}h_{SR}X[i-1] + Z[i] + \alpha h_{RD}Z_R[i-1].
\end{equation}
The capacity of the Gaussian ISI channel is given by
\begin{equation} \label{AF_rate}
R_{AF}= \max_{S(\omega)} \frac{1}{2\pi} \int_{0}^{2\pi} \frac{1}{2} \log \left( 1 + S(\omega)\frac{|H(\omega)|^2}{N(\omega)}\right)d \omega,
\end{equation}
subject to
\begin{equation}
\frac{1}{2\pi}\int_{0}^{2\pi}S(\omega)d\omega \le 1, \;\; \textrm{and}
\;\;  S(\omega) \ge 0, \quad 0 \le \omega \le 2\pi,
\end{equation}
where $N(\omega) = 1 + \alpha^2 h_{RD}^2 + 2\rho_z\alpha h_{RD} \cos(\omega)$
is the power spectrum density of the noise, and $H(\omega) = h_{SD} + \alpha h_{RD}h_{SR}e^{j\omega}$
is the Fourier transform of the channel coefficients, and $S(\omega) = \left(\lambda - \frac{N(\omega)}{|H(\omega)|^2} \right)^+$
is the water-filling power allocation over the frequencies.

Consider again the case of $\rho_z = 1$ and the asymptotic regime of
(\ref{extreme_scenario}), i.e.  $h_{RD}^2 \gg h_{SR}^2 \gg h_{SD}^2 \gg 1$.
In this high signal-to-noise ratio regime, it
is easy to verify that 
the water-filling power spectrum converges to an equal power allocation, i.e.,
$S(\omega) = 1$, $0\le \omega \le 2\pi$. Substituting $N(\omega)$, $H(\omega)$
and $S(\omega) =1$ into (\ref{AF_rate}) and using table of integrals, after some algebra, it is possible to show that 
\begin{eqnarray*}
R_{AF} \le \frac{1}{2} \log (2 + h_{SR}^2 + h_{SD}^2).
\end{eqnarray*}
Comparing the above with the cut-set bound, we see that
\begin{eqnarray}
\overline{C} - R_{AF} &\ge& \hf \log
	\left(\frac{1 + h_{SD}^2 + h_{RD}^2 + 2h_{SD}h_{RD}}
	{2 + h_{SR}^2 + h_{SD}^2} \right)  \nonumber \\
	&\rightarrow& \hf \log \left(\frac{h_{RD}^2}{h_{SR}^2} \right)
\end{eqnarray}
in the asymptotic regime of (\ref{extreme_scenario}), which is unbounded.

\section{Numerical Simulation}

\begin{figure} [t]
\centering
\includegraphics[width=3.1in, height=2.4in]{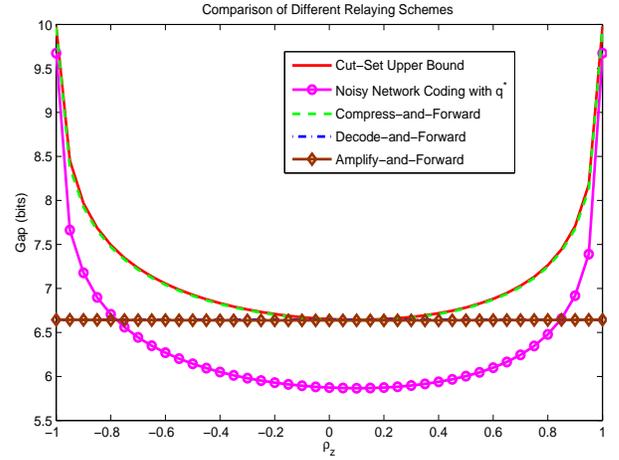}
\caption{Comparison of different relaying schemes for
$h_{SD}^2 = 20$dB, $h_{SR}^2=40$dB and $h_{RD}^2=60$dB}
\label{Fig_comparison}
\end{figure}

\begin{figure} [t]
\centering
\includegraphics[width=3.15in, height=2.4in]{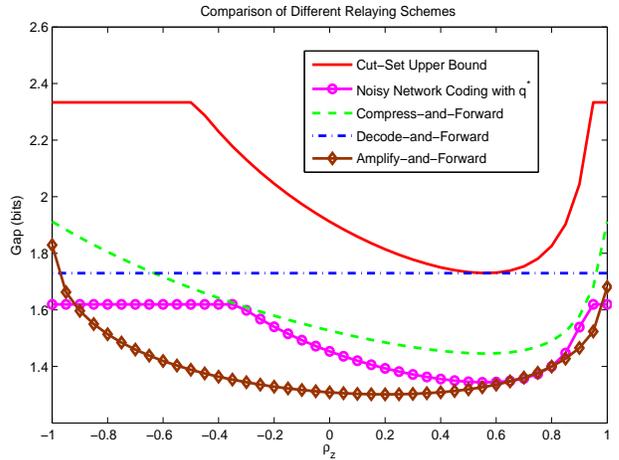}
\caption{Comparison of different relaying schemes for
$h_{SD}^2 = 5$dB, $h_{SR}^2=10$dB, and $h_{RD}^2=10$dB}
\label{Fig_comparison_2}
\end{figure}

This section numerically compares the cut-set upper bound and the achievable rates of different relaying schemes. Here, the noisy network coding rate is computed with $\q^*=2(1-\rho_z^2)$. We consider two examples: Fig.~\ref{Fig_comparison}
shows the case for $h_{SD}^2 = 20$dB, $h_{SR}^2=40$dB and
$h_{RD}^2=60$dB, corresponding to an extreme scenario of $h_{RD}^2 \gg h_{SR}^2 \gg h_{SD}^2 \gg 1$. Fig.~\ref{Fig_comparison_2} shows the case
for $h_{SD}^2 = 5$dB, $h_{SR}^2=10$dB, and $h_{RD}^2=10$dB. It is clear that in both cases, compress-and-forward is always better than the noisy network coding scheme with the specific $\q^*$, and both are within a constant gap to the cut-set upper bound for all values of $\rho_z$.

The decode-and-forward rate is always independent of $\rho_z$.
In the asymptotic regime as shown in Fig.~\ref{Fig_comparison}, the single-tap amplify-and-forward rate is almost independent of $\rho_z$ as well, and it coincides with the decode-and-forward rate. Both can have an unbounded gap to the cut-set bound as $h^2_{RD} \ra \infty$ and $\rho_z \ra \pm 1$.  Compress-and-forward, on the other hand, closely tracks the cut-set bound.  (Note that the above observations are not true in the  non-asymptotic SNR regime as shown in Fig.~\ref{Fig_comparison_2}.) The noisy-network-coding scheme, although not as good as compress-and-forward, nevertheless is always within a constant gap to the cut-set bound.

It is interesting to see that the noisy-network-coding rate resembles the shape of the cut-set upper bound as shown in both Fig.~\ref{Fig_comparison} and Fig.~\ref{Fig_comparison_2}.
It is also interesting to note that the decode-and-forward curve touches the cut-set bound at a particular value of $\rho_z$. This is because at this value of $\rho_z$, the relay channel becomes degraded \cite[Theorem~1]{Cui_correlated}.



\section{Conclusion}
This paper investigates different relaying strategies for the three-node Gaussian relay channel with correlated noises. It is shown that both the proposed correlation-aware noisy network coding scheme and the conventional compress-and-forward relaying scheme can achieve to within a constant gap to the capacity, while the decode-and-forward scheme and the single-tap amplify-and-forward scheme cannot.

\bibliographystyle{IEEEtran}
\bibliography{IEEEabrv,../ref/main}

\end{document}